# The Classical-to-Quantum Crossover in strain-induced ferroelectric transition in SrTiO$_3$ membranes

Jiarui Li[1], Yonghun Lee[1,2], Yongseong Choi[3], Jong-Woo Kim[3], Paul Thompson[4], Kevin J. Crust[1,5], Ruijuan Xu[6], Harold Y. Hwang[1,2*], Philip J. Ryan[3*], Wei-Sheng Lee[1*]

[1]Stanford Institute for Materials and Energy Sciences, SLAC National Accelerator Laboratory, Menlo Park, CA, USA
[2]Department of Applied Physics, Stanford University, Stanford, CA, USA
[3]Advanced Photon Source, Argonne National Laboratory, Lemont, IL, USA
[4]Oliver Lodge Laboratory, Department of Physics, University of Liverpool, Liverpool, L69 7ZE UK
[5]Department of Physics, Stanford University, Stanford, CA 94305, USA
[6]Department of Materials Science and Engineering, North Carolina State University, Raleigh, NC 27606, USA

**Mechanical strain presents an effective control over symmetry-breaking phase transitions. In quantum paralelectric SrTiO$_3$, strain can induce the ferroelectric transition via modification of local Ti potential landscape. However, brittle bulk materials can only withstand limited strain range (~0.1%). Taking advantage of nanoscopically-thin freestanding membranes, we demonstrated in-situ strain-induced reversible ferroelectric transition in a single freestanding SrTiO$_3$ membranes. We measure the ferroelectric order by detecting the local anisotropy of the Ti 3$d$ orbital using X-ray linear dichroism at the Ti-$K$ pre-edge, while the strain is determined by X-ray diffraction. With reduced thickness, the SrTiO$_3$ membranes remain elastic with >1% tensile strain cycles. A robust displacive ferroelectricity appears beyond a temperature-dependent critical strain. Interestingly, we discover a crossover from a classical ferroelectric transition to a quantum regime at low temperatures, which enhances strain-induced ferroelectricity. Our results offer a new opportunities to strain engineer functional properties in low dimensional quantum materials and provide new insights into the role of the ferroelectric fluctuations in quantum paralelectric SrTiO$_3$.**

Controlling the properties of emergent quantum phases are of fundamental importance for modern nanotechnology. Strain control provides a direct manipulation over the atomic spacing, resulting effective tuning of quantum phases through the intertwined structural, electronic and magnetic degrees of freedom in quantum materials. While strain has been used to tune properties of a variety of bulk materials[1–7], its application to bulk crystals is limited due to their low tensile strain tolerance, typically around 0.1%[8]. The advent of low dimensional quantum material membranes, such as 2D van der Waals materials and oxide membranes, has enabled new opportunities for mechanical manipulation of quantum states of matter[4,9]. Studies have demonstrated that these low dimensional materials, especially oxide freestanding nano-membranes, can sustain a degree of strain well beyond their bulk counterparts, allowing the exploration of coupling between lattice, charge, spin, and orbital degrees of freedom in as yet unexplored strain regimes[10–15].

Since ferroelectricity arises from asymmetric ion displacement, strain provides an effective tuning on the local potential energy and plays a central role in the control of ferroelectricity[10,16–18]. In this study, we focus on $SrTiO_3$, known as a "quantum paraelectric", whose ground state is believed to lie near a ferroelectric quantum critical point. Despite a dramatic increase of dielectric constant when approaching the zero temperature, strong quantum fluctuations destabilizes the spontaneous development of long-range static ferroelectric order[19,20]. Various non-thermal tuning parameters, such as oxygen isotope exchange[21], epitaxial strain[16,17], dopings, and dislocations[22,23], can overcome the fluctuations and stabilize the ferroelectric order in $SrTiO_3$. While widely acknowledged as a displacive transition[24–26], several studies have reported randomly oriented polar nanodomains preceding the ferroelectric order in $SrTiO_3$[27–32], illustrating a relaxor ferroelectric nature where the long-range order arises from the alignment of preformed ferroelectric nanodomains (Fig. 1a). The precise nature of the non-thermally induced ferroelectric transition remains elusive. To unravel the role of the thermal and quantum fluctuations, as well as the displacive and relaxor characteristics, a comprehensive investigation of the $SrTiO_3$ ferroelectric phase diagram with consistent sample heterogeneity is essential. Previous studies on induced ferroelectricity, however, typically involve multiple samples[10,16,17,21–23], which unavoidably introduces sample variations and hinders a deeper understanding of the transition and limits the technology application via strain tuning.

In this study, we leverage the opportunity offered by freestanding membranes of $SrTiO_3$ and implemented an *in-situ* strain apparatus capable of applying large uniaxial strains to low-dimensional membrane materials for multi-modal X-ray measurements. This new capability allows for the continuous and reversible manipulation of the lattice parameter of freestanding SrTiO$_3$ membrane, and accessing the ferroelectric phase transition to study the regime across phases of thermal and quantum-driven fluctuations. The strain response of the lattice constants at various cryogenic temperatures is monitored by X-ray diffraction (XRD) and the induced ferroelectric order parameter is tracked by linear dichroic X-ray absorption (XLD). By quantitatively correlating the Ti-$K$ pre-edge XLD spectral feature with ferroelectric polarization moments across a wide strain and temperature range in a single SrTiO$_3$ nanomembrane, we present evidence of a temperature-dependent crossover from a high-temperature classical phase transition into a distinct low-temperature quantum paraelectric regime where quantum fluctuations contribute significantly.

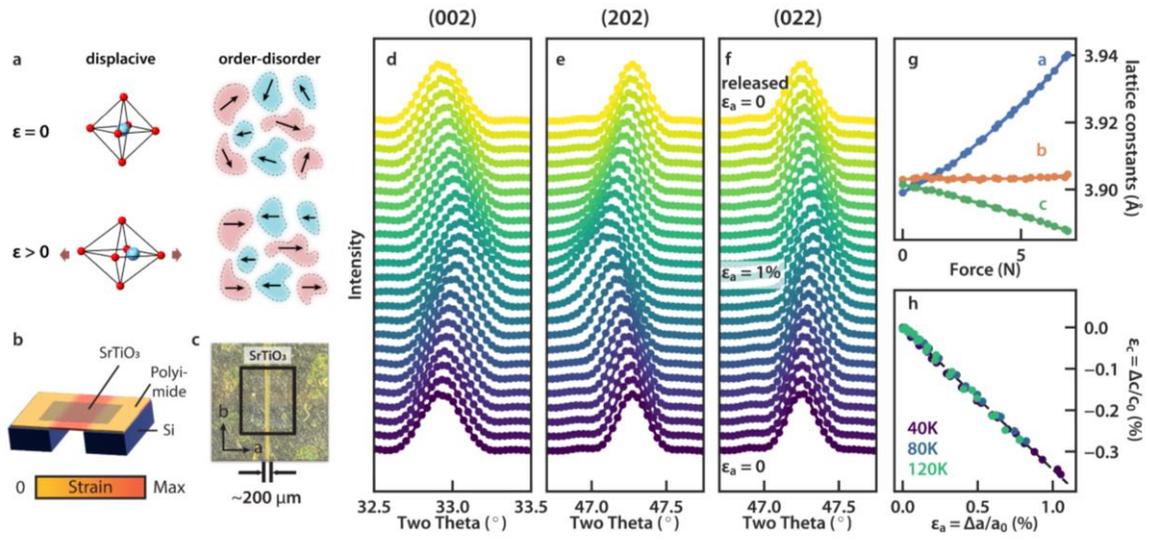

*Figure 1. Straining SrTiO$_3$ membrane. a, Two scenarios of the strain-induced ferroelectric transition: displacive (left) and order-to-disorder (right). b, Schematic of the uniaxial strain applied onto SrTiO$_3$ sample supported by polyimide films. The device is suspended over a hundred-μm gap defined by two cleaved silicon substrates. The schematic drawing is not to scale. c, Image of the SrTiO$_3$ sample near the gap area. d-f, Single crystal X-ray diffraction of three Bragg peaks [(002), (202), (022), in pseudocubic notation] during the tensile strain and release process at 40 K. g, Refined lattice parameters of the strained*

*SrTiO$_3$ with different applied forces along crystallographic **a** direction via piezo stacks. **h,** The relationship between the lattice strain of two relevant lattice parameters, namely the length along the strain $\varepsilon_a = \Delta a/a_0$ and out-of-plane $\varepsilon_c = \Delta c/c_0$, for selected temperatures, where $a_0$ and $c_0$ are the lattice constants at zero strain.*

## Results

**Reversible strain control**

We first demonstrate the reversible tuning of the lattice over 1% of tensile strain in a 20 nm thick SrTiO$_3$ membrane on the uniaxial strain setup (Fig. 1b,c). The uniaxial strain direction is parallel to the pseudocubic **a** axis, with the **b** axis in-plane orthogonal to the strain direction (along the edges of the silicon gap), and **c** is the out-of-plane direction of the SrTiO$_3$ membrane. The high-resolution X-ray diffraction of the SrTiO$_3$ membranes at 40 K and different strain values are shown in Figure 1d-f. The (002), (202), and (022) Bragg peaks exhibit systematic shifts as the elastic strain is applied and released, indicating a reversible tuning in the lattice constants. Extensile strain generates an increasing *a*, and a decreasing *c*, while *b* remains unchanged due to the constraint of the silicon substrates (Fig. 1g). This allows us to determine the Poisson's ratio by calculating the linear dependence of the strains $\nu = -\varepsilon_c/\varepsilon_a$, where the strain is defined by $\varepsilon_a = \Delta a/a_0$, and $a_0$ is the lattice constant at zero strain. We found $\nu = 0.34$ with little temperature dependence (Fig. 1h), which is notably larger than that of the bulk ($\nu = 0.22$)[33]. The width of the in-plane (202), (022) Bragg peaks also increase upon strain, which could indicate the increase number of dislocations that destroys the in-plane lattice coherence. The peak width recovers when strain is released.

While similar lattice responses were observed in SrTiO$_3$ membranes with varying thicknesses and temperatures, we found that the elastic strain limit increased fivefold upon decreasing membrane thickness from 50 nm to 10 nm[15].

**Strain-induced ferroelectric order**

Having established the elastic lattice response, we now demonstrate that the uniaxial strain can also induce a reversible paraelectric to ferroelectric transition in the SrTiO$_3$ membrane. X-ray absorption spectroscopy (XAS) provides a sensitive probe of local symmetry at a specific atomic site. The XAS near the Ti-*K* edge results from the dipole transition from Ti 1*s* to Ti 4*p* orbitals (around 4.982 keV in Fig. 2a), with mixtures of weaker transitions between Ti 1*s* and Ti 3*d* ($e_g$ and $t_{2g}$) orbitals in the pre-edge region, where the latter is dipole-forbidden but quadrupole-allowed in the paraelectric states[34,35]. Upon transition into the ferroelectric state that breaks inversion symmetry, the Ti 1*s* to Ti 3*d* $e_g$ transition becomes dipole allowed in XAS when the X-ray polarization is parallel to the Ti atom displacement (**E**//***a***, throughout this manuscript)[34,35]. Figure 2b shows the strain-induced ferroelectric transition by showing the change of the Ti-*K* XAS spectra at a series of strain values, taken at 20 K with grazing incident geometry **E**//***a***. We observe a gradual enhancement of the $e_g$ XAS pre-edge peak centered at 4970.8 eV with increasing strain indicating a strain-induced change in the local environment of the Ti ions. The change in XAS is only present when the X-ray polarization is along the tensile strain direction (**E**//***a***), as the induced ferroelectric polarization is parallel to the tensile strain (Fig. 2c). Moreover, the independence of the **E**//***c*** XAS spectra to the tensile strain suggest no change of ferroelectric polarization along ***c***.

This observation appears to be inconsistent with the order-disorder ferroelectric transition scenario, where the microscopic preformed ferroelectric domains are randomly oriented along different directions in the unstrained disordered phase (Fig. 1a)[30]. Uniaxial strain aligns Ti displacement along ***a*** and perpendicular to ***c*** (Fig. 1a). One would expect suppression of the $e_g$ peak intensity in **E**//***c*** channel upon tensile straining ($\varepsilon_a$) if the transition is order-disorder in nature[15]. The insensitivity of **E**//***c*** spectra with respect to strain puts an upper bound on the preformed polar nano domain with maximum Ti displacement of $\delta_{Ti}$ < 0.7 pm, which is an order smaller than the thermal vibration at 50 K[36]. Therefore, our results strongly support a displacive nature of the strain-induced ferroelectric transition in SrTiO$_3$.

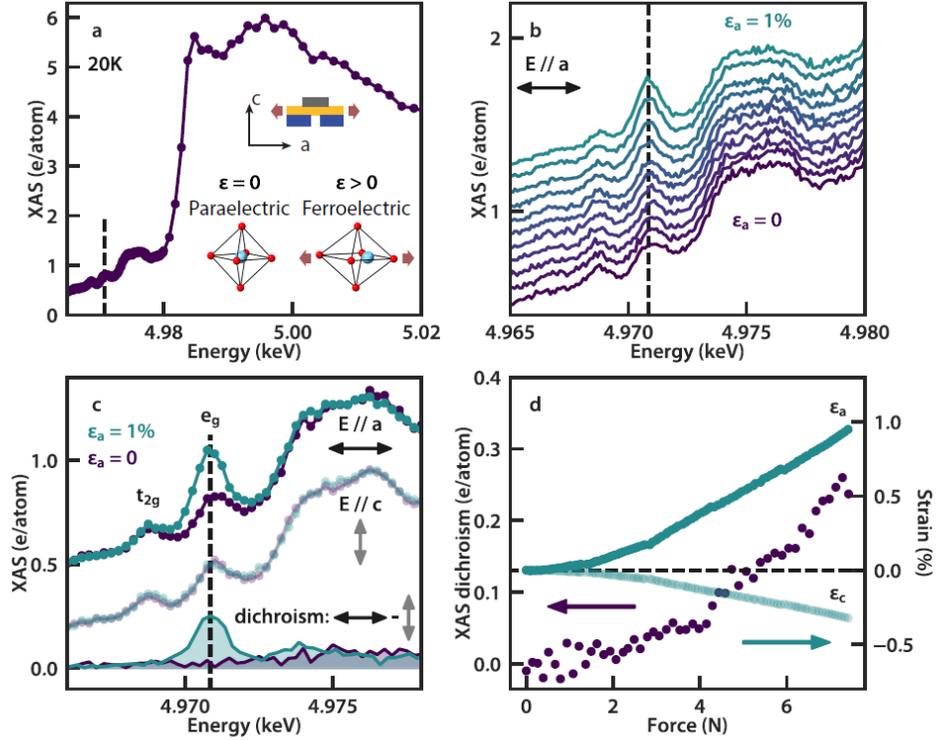

***Figure 2. Strain-induced ferroelectric state at 20 K. a,*** *XAS of unstrained SrTiO₃ membrane across Ti K-edge. The inset shows the schematics of the strain geometry and the strain-induced Ti displacement.* ***b,*** *The strain evolution of the XAS in the pre-edge regime. The strained spectra correspond to the ferroelectric state.* ***c,*** *XAS pre-edge feature for X-ray polarizations along different crystallographic axis (**E//a** and **E//c**). The first two peaks in the unstrained spectra are quadrupole transitions of the Ti 1s electrons to the Ti 3d-derived $t_{2g}$ and $e_g$ orbitals. The non-trivial linear dichroic signal at $e_g$ pre-edge peak (at 4970.8 eV, vertical dashed line) emerges when SrTiO₃ is strained.* ***d,*** *The evolution of the linear dichroic signal (purple dots) at $e_g$ pre-edge peak (vertical dashed line in a-c) and the strain in **a** and **c** lattice parameter, denoted as $\varepsilon_a$ and $\varepsilon_c$, respectively, upon applying uniaxial strain along **a**.*

As a quantitative measure of the ferroelectric order, XAS linear dichroism spectra (XLD) is the difference between **E//a** and **E//c**, as shown in Fig 2c. The strained XLD spectra shows spectral weight at the $e_g$ position which is absent in the unstrained state. This intensity is proportional to the square of the ferroelectric order parameter: XLD $\propto |P^2|$[35]. Compared with previous XAS and DFT calculations[34], the maximum polarization is estimated to be around P = 14.0 μC/cm² with displacements for Ti and averaged O atoms are estimated to be around $\delta_{Ti}$ = 2.01 pm and $\delta_O$ = 7.30 pm[15].

We thus tracked this strain-dependent parameter by monitoring the XLD $e_g$ intensity as we increased the strain force from zero. Figure 2d shows XLD intensity increasing monotonically, indicating a continuous growth of ferroelectric order under tensile strain. By monitoring the change of the *a*, *c* lattice constant during the strain loop, the calculated the *a*, *c* axis strain $\varepsilon_{a,c}$ is overlaid in figure 2d. By relating the lattice response to XLD spectra at the same strain state, a strain-dependent ferroelectric order parameter $P(\varepsilon_a)$ is established. Figure 3a summarizes the strain response at different temperatures up to 100 K. The presented XLD data were obtained with simultaneous XRD measurements to track the lattice response upon straining. At 20 K, a clear dichroic signature emerges at finite tensile strain, suggesting the appearance of ferroelectric order at a small tensile strain value. Whereas at higher temperatures, ferroelectric order appears only beyond a critical strain value $\varepsilon_0(T)$, which increases with rising temperature. Importantly, at each temperature, the strain-induced ferroelectric order evolves smoothly near $\varepsilon_0$ without a discontinuous jump, signifying a second-order phase transition driven by strain[37].

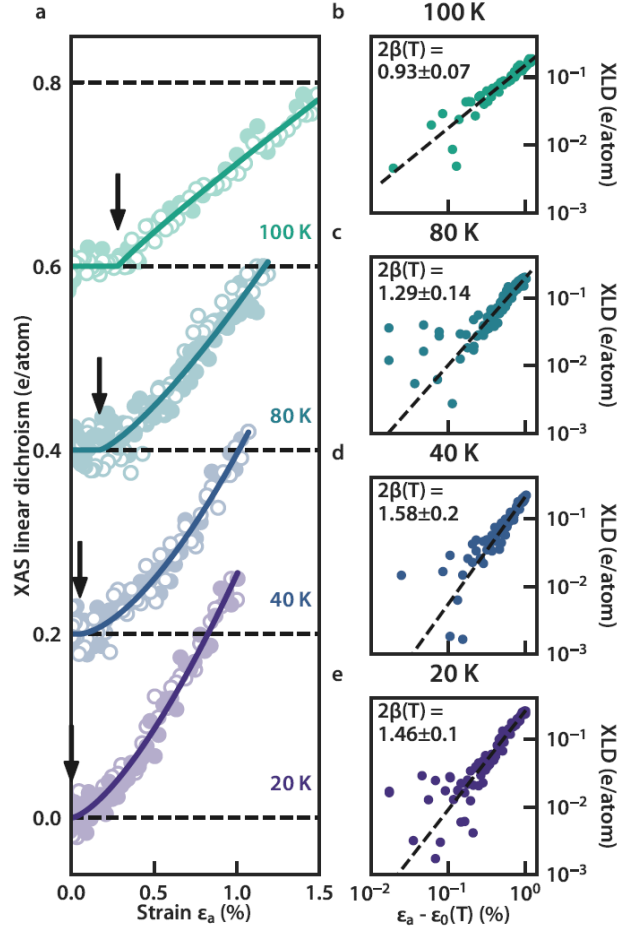

*Figure 3. **Temperature/strain-dependent ferroelectric order. a,** Strain-dependent XAS linear dichroic (XLD) signal at different temperatures. The filled (open) circle marks the increasing (decreasing) strain sweep directions. The solid line is the fit to a power law behavior as in **b-e**. Data is offset vertically for clarity. **b-e,** Fits of power law dependence of ferroelectric order to the lattice strain at different temperatures.*

**Dramatic change in the critical exponent β upon cooling.**

Intriguingly, we find that the XLD signal vs. strain displays distinct curvatures at different temperatures. Near a critical phase boundary, one expects physical quantities to exhibit critical behavior following the order parameter $P(\varepsilon) \propto [\varepsilon_a - \varepsilon_0(T)]^\beta$, where $\beta$ is the order parameter critical exponent. The change in curvature suggests a non-universal critical exponent $\beta$. Fitting the strain-dependent XLD to the power law behavior XLD $\propto |P^2| \propto [\varepsilon_a - \varepsilon_0(T)]^{2\beta}$ shows a temperature-dependent critical exponent $2\beta(T)$, around 0.93 at 100 K, while increasing to around 1.5 at below 40 K (Figure 3b-e).

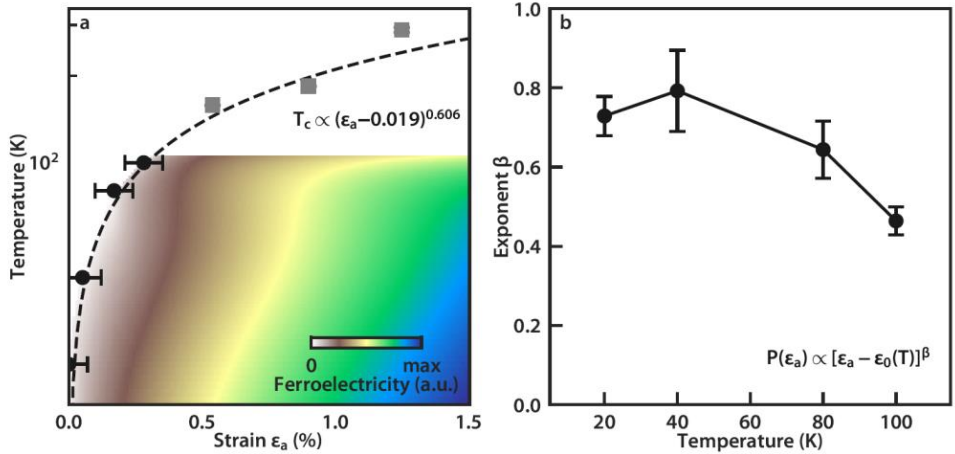

*Figure 4. **Paraelectric-ferroelectric phase boundary and phase diagram. a,** Temperature-strain phase diagram of the paraelectric-ferroelectric phase. The dash-line is the fitted phase boundary to the power law. In the ferroelectric side of the phase diagram, the strength of the ferroelectric order parameter is color-coded. The grey data points are reproduced from Ref. [10]. **b,** Temperature dependence of the order parameter critical exponent β.*

Figure 4 summarizes the temperature-strain phase diagram for the paraelectric-ferroelectric phase transition of $SrTiO_3$ membrane. This phase boundary is consistent with previous studies that reported the strain-induced ferroelectricity in $SrTiO_3$ thin films and single crystals, where the phase boundary terminates at $T=0$ quantum critical point (QCP) near zero strain value[10,38]. From the paraelectric-ferroelectric phase boundary, we determine the critical scaling behavior to be a power law, where $T_c \propto [\varepsilon_a - \varepsilon_0(T=0)]^{1/\Delta}$, with the extrapolated quantum critical strain being $\varepsilon_0(T=0) = 0.019\% \pm 0.016\%$, and $1/\Delta = 0.606 \pm 0.13$ (Fig. 4a). Figure 4b summarizes the order parameter critical exponent $\beta$ across the paraelectric-ferroelectric phase boundary. At high temperatures, $\beta \sim 0.5$, which is consistent with the classical ferroelectric transition predicted in a mean-field theory, assuming a linear-quadratic strain-ferroelectric order coupling in the Landau free energy[15]. Interestingly, with decreasing temperature, the critical exponent $\beta$ substantially increases upon cooling below 100 K, suggesting a change in the nature of the phase transition.

## Discussion

$SrTiO_3$ has been argued to be a quantum paraelectric near a displacive ferroelectric QCP where the paraelectric-ferroelectric phase boundary terminates at zero temperature[10,38]. Thus, as the temperature decreases, thermal fluctuations diminish, and quantum fluctuations should play a dominant role. Our observation of temperature dependent β strongly supports the existence of the QCP. Furthermore, our results indicate that the influence of the quantum criticality sets in below ~ 100 K, consistent with the deviation from the classical Curie Weiss behavior in paraelectric dielectric susceptibility[19]. The critical exponent $\beta$ saturates at around ~50K, also aligned with temperature range where the dielectric constant exhibits a non-classical $T^2$ dependence[39,40]. Apparently, to quantitatively understand the temperature variation of $\beta$ in our data, a quantum correction is needed to account for the deviation from the classical critical behavior. In the dimensionality cross-over theory, quantum criticality near a QCP increases the effective dimensionality to $d + z$, where $d$ is the original diemsnionality and $z$ is the dynamical exponent[39,41]. Such that the quantum critical behavior is analogous to a classical transition, but in a higher dimension. Our observation of an increasing $\beta$ in the proximity of the quantum critical reqime at low temperature appears to be qualitatively consistent with the increase of an effective dimensionality. However, since strain-ferroelectric order coupling in $SrTiO_3$ is linear-quadratic, it cannot be simply described by the Ising model with a transverse conjugate field that is commonly used as a model system for QCP theories. Early theory based on ferroelectric systems suggested a logarithmic correction necessary for the quantum fluctuation effect under isotropic pressure[25,42]. Whether a similar correction applies to our observed uniaxial strain-induced ferroelectric transition remains a compelling theoretical question.

Could the presence of spatial inhomogeneity at low temperatures account for the observed temperature dependence of $β$? Previous studies have suggested that frozen-in disorders at low temperatures, such as sample or strain inhomogeneity, could lead to a distribution of critical strain value $ε_0$, smearing out the ferroelectric transition and increasing $β$[43]. We argue that this scenario is unlikely to explain our data. Inhomogeneity often leads to history-dependent ferroelectric responses, but our strain-dependent ferroelectric order parameter shows no hysteresis within experimental error (Fig. 3a). Moreover, we show the distribution of the lattice strain $ε_a$ across the gap region in the Supplementary Fig. 7[15]. In the center of the gap, the strain is homogenous with a strain variation $Δε_a$ smaller than 0.09%. Such small inhomogeneity is unlikely to induce a large shift in the order parameter critical exponent $β$.

## Conclusion and outlook

To conclude, we have demonstrated reversible in-situ strain engineering of a single $SrTiO_3$ membrane device across paraelectric-to-ferroelectric transition. Our results suggest that the strain-induced ferroelectric transition is displacive in nature with no preformed nano polar domains. Using multi-modal X-ray techniques, we quantified the ferroelectric moment from the Ti-$K$ pre-edge features and chart out the order parameter in an extended strain-temperature phase diagram. The temperature dependent strain-order parameter critical exponent $β(T)$ indicates a classical-to-quantum crossover between two distinct regimes.

In a broader context, the reversible strain-tuning appears to be a gerenic property of quantum material nano-membrane, as shown in Supplementary Fig. 4, in which the reversibility is demonstrated on two other strained perovskite oxide membranes, $La_{0.7}Ca_{0.3}MnO_3$ and $LaNiO_3$. Unlike previous works in bulk crystals, which often have high failure rate at higher strain values, laminating nano-membranes on stretchable polymer platform offers additional support for strain engineering while maintaining structural integrity. This open new avenues for controlling and probing exotic quantum states of matter through robust lattice control in a wide class of quantum materials. For example, we envision continuous and reversible control of the electronic structures[2,44] and complex magnetism[45] over an extended temperature-strain phase diagram, allowing precise access of critical phase boundaries of competing orders and remaining compatible to a broader range of characterizations. These developments also promise novel microelectromechanical devices that can harness the extreme properties of quantum materials, such as superconductivity, magnetism, resistive switching, and ferroelectricity, through nano-mechanical control. By providing flexible, lightweight, and highly durable quantum material platforms, this reversible strain-tuning approach has broad implications for soft quantum electronics, including wearable devices, implantable devices, and advanced sensors.

**Methods**

**Thin film growth and membrane fabrication.** The free-standing crystalline SrTiO$_3$ was synthesized by pulsed layer deposition with designed thickness with a 15.6 nm Sr$_2$CaAl$_2$O$_6$ sacrificial buffer layer on (001)-oriented single crystalline SrTiO$_3$ substrates. To improve the quality of resulting SrTiO$_3$ membranes and minimize fractures, the sacrificial buffer layer was chosen to closely match the SrTiO$_3$ lattice parameters. Before the growth, the (001)-oriented single crystalline SrTiO$_3$ substrate was pre-annealed at 900°C and oxygen partial pressure of 5×10$^{-6}$ Torr to achieve sharp step-and-terrace TiO$_2$-terminated surfaces. A 40-unit cell of Sr$_2$CaAl$_2$O$_6$ sacrificial buffer layer was first grown by pulsed layer deposition using a 248 nm KrF excimer laser with 2.1 J/cm$^{-2}$ laser fluence, 3.1 mm$^2$ laser spot size on the target, and substrate temperature of 710°C. A subsequent SrTiO$_3$ with target thickness was grown with 0.89 J/cm$^{-2}$ laser fluence, 3.2 mm$^2$ laser spot size, and substrate temperature of 700°C. The growth process was monitored via in-situ reflection high-energy electron diffraction (RHEED).

The thin, as-grown heterostructure was first coated with PMMA and placed in deionized water until the Sr$_2$CaAl$_2$O$_6$ sacrificial layer was fully dissolved. The released membrane was transferred onto a polyimide film. Prior to the transfer, the polyimide film was treated with oxygen plasma to obtain an activated surface. The samples were annealed at 110 °C for 10 minutes to enhance SrTiO$_3$-polyimide bonding. Finally, the PMMA was removed by acetone.

To be able to apply large strain to SrTiO$_3$ membrane, two pieces of cleaved silicon substrates were mounted onto a titanium flexure plate (Razorbill Instruments), with a well-defined hundred-µm parallel gap in between. The polyimide-supported SrTiO$_3$ was mounted across the gap using Angstrom Bond epoxy. The whole sample assembly is then mounted to an FC150 strain cell from Razorbill Instruments. The SrTiO$_3$ (100) pseudocubic crystallographic axis was chosen to align with the uniaxial strain direction. We use pseudocubic notation throughout the study. The schematic drawing and photo of the device are shown in Fig. 1b,c and the Supplementary Figure 6[15]. The small gap defined by two cleaved silicon substrates can provide a maximum of more than 20% uniform tensile strain for the material suspended above the gap, given the sample does not fail.

**X-ray measurements**: X-ray diffraction (XRD) and absorption measurements were performed at the 6-ID-B and 4-ID-D endstation of the Advanced Photon Source, Argonne National Laboratory. Part of the measurements were conducted at XMaS beamtline. The FC100 strain cell was mounted into an Advanced Research Systems Displex closed-cycled cryostat with a base temperature T ≈ 20 K. The X-ray beam spot was kept smaller than the gap size in any geometry to ensure only strained sample region is probed. XRD of (002), (202), (022) Bragg peaks were performed at photon energy of 11.212 keV with Gaussian fits to determine the pseudocubic lattice constants.

XAS spectra across the Ti-*K* edge resonance (4.97 keV) were recorded in partial fluorescence yield mode in a quasi-backscattering geometry with a Hitachi Vortex detector. At each incident energy, the polarization was rapidly switched between linear polarization states using diamond phase plates in the sequence of LV/LH/LH/LV in order to determine the linear dichroism spectrum with a high signal-to-noise ratio. The X-ray was incident at 1 degree grazing angle along the gap, where LH/LV polarization corresponds to polarization vector along *a*/*c* pseudocubic axis of SrTiO$_3$, respectively. All XAS/XLD spectra were normalized to the Ti-*K* main edge jump up to 5.02 keV with the tabulated value[46]. Back-to-back to XAS measurements, strain dependent *c* lattice constant was tracked from (002) Bragg peak XRD at photon energy of 5.5 keV. The compressive strain along **c** axis was converted to **a** axis tensile strain using the Poisson's ratio $\varepsilon_a = -\varepsilon_c/\nu$.

**Acknowledgments**

We gratefully thank Alexander Balatsky, Hennadii Yerzhakov, Han Zhaoyu, Steve Kivelson, Yiming Wu, Wei-Cheng Lee, Peter Abbamonte, Mark Dean, Makoto Hashimoto for the stimulating and fruitful discussions. This work was supported by the U. S. Department of Energy, Office of Basic Energy Sciences, Division of Materials Sciences and Engineering (Contract No. DE-AC02-76SF00515), and the Laboratory

Directed Research and Development program at SLAC National Accelerator Laboratory (early stage development). The work performed at the Advanced Photon Source was supported by the US Department of Energy, Office of Science, Office of Basic Energy Sciences under contract no. DE-AC02-06CH11357. XMaS is a UK national research facility supported by EPSRC. We are grateful to all the beamline team staff for their support.

**Author Contributions**

J. L., H. Y. H., P. R. and W. S. L. conceived the project. J. L. and P. R. designed the experiment. J. L. and Y. L. prepared and characterized the membrane samples with K. J. C. and R. X.'s contribution. J. L., Y. C., J. K., P. T. and P. R. conducted the experiment. J. L., P. R. and W. S. L. analyzed the data and wrote the manuscript with critical input from all authors.